\documentclass{aa}  

\usepackage{graphicx}
\usepackage{txfonts}
\begin{document}


\title{$ThermoONet$ - a deep learning-based small body thermophysical network: applications to modelling water activity of comets}
\author{{Shunjing Zhao\inst{1,2,3}},
        {Xian Shi\inst{3}},
        {Hanlun Lei\inst{1,2}}}
\institute{School of Astronomy and Space Science, Nanjing University, Nanjing 210023, China\label{inst1}\and
Key Laboratory of Modern Astronomy and Astrophysics in Ministry of Education, Nanjing University, Nanjing 210023, China\label{inst2}\and
Shanghai Astronomical Observatory, Chinese Academy of Sciences, Shanghai 200030, China\label{inst3}\\
\email{shi@shao.ac.cn}\\
\email{leihl@nju.edu.cn}}

\date{accepted April, 2025}

\abstract
{Cometary activity is a compelling subject of study, with thermophysical models playing a pivotal role in its understanding. However, traditional numerical solutions for small body thermophysical models are computationally intensive, posing challenges for investigations requiring high-resolution or repetitive modeling. To address this limitation, we employed a machine learning approach to develop $ThermoONet$ — a neural network designed to predict the temperature and water ice sublimation flux of comets. Performance evaluations indicate that $ThermoONet$ achieves a low average error in subsurface temperature of approximately 2$\%$ relative to the numerical simulation, while reducing computational time by nearly six orders of magnitude. We applied $ThermoONet$ to model the water activity of comets 67P/Churyumov-Gerasimenko and 21P/Giacobini-Zinner. By successfully fitting the water production rate curves of these comets, as obtained by the Rosetta mission and the SOHO telescope, respectively, we demonstrate the network's effectiveness and efficiency. Furthermore, when combined with a global optimization algorithm, $ThermoONet$ proves capable of retrieving the physical properties of target bodies.}

\titlerunning{ThermoONet}
\authorrunning{Zhao et al.}

\maketitle

\section{Introduction}\label{sec1:intro}
Comets are active small solar system bodies. When approaching the Sun, their nuclei get heated up, and volatile ices therewithin sublimate, dragging dust particles alongside to form detectable coma and tails \citep{Whipple1950}. The activity of comets is an intriguing topic, as it provides insights into the formation and evolution of comets, such as the cyclical variations in the relative abundance of water ice keeping the comet ‘alive’ \citep{Sanctis2015,Shi2018}, heat transport and ice sublimation reflecting properties of the upper-most layer of cometary nucleus \citep{Gulkis2015}, seasonal effect of ice sublimation inducing various topographical changes \citep{El-Maarry2015,Tang2024}. It can also help us to understand the possibility for comets to have delivered water to Earth \citep{Mantdt2024}.

Within $\sim$ 3 au, water vapor is the predominant species in outgassing \citep{Hassig2015,Lauter2020}. The water production rate and the dust production rate are primary indicators for characterizing cometary activity and its variations. The modelling of water and dust production requires simulating the thermophysical processes, taking into account the orbit and the shape of the nucleus, and its various physical properties. There are variations in the thermophysical models for different aims \citep{Watson1963,Cowan1979,Weissman1981}. They have been applied in quantitative analysis of the impact of parameters on water ice activity \citep{Marshall2019}, as well as investigating mechanisms behind the disruption of a cometary nucleus \citep{Steckloff2015}. For interpreting observational data, more realistic thermophysical models are required, such as the inversion of 67P's physical properties from non-gravitational force caused by outgassing \citep{Davidsson2005}, the evolution of morphology of the nucleus influenced by the insolation-induced activity \citep{Keller2015, Groussin2025}, interpretation of the measured water production \citep{Hu2017,Blum2017,Attree2019}, and the explanation of distinct activity phenomena such as the short-lived outbursts from fractured terrains \citep{Skorov2016}. To further improve model fidelity, it is necessary to consider additional complex physical mechanisms, such as the behavior of volatile transport in the nucleus given rise by its porosity, the deposition of gas on dust particles and so on \citep{Sanctis1999,Davidsson2002,Gortsas2011,Gonzalez2014,Hu2021}.

Traditionally, thermophysical models are realized by numerical treatment based on the finite-difference technique, such as the Crank-Nicolson method utilized by \citet{Hu2017}. However, it usually requires relatively high computation cost due to the iterative calculations performed on a large time-space grids, leads to limited efficiency in related studies, posing challenges for practical implementation. Firstly, the inverse extraction of physical parameters automatically from observational data is a difficult problem, typically, the forward trial-and-error method is utilized \citep{Keller2015,Hu2017,Attree2019}. Besides, \citet{Combi2019,Combi2021b,Combi2023} are continuously updating the water production rate curves of numerous comets, directly solving the thermal models through numerical simulation to analyze these data is clearly a highly challenging task. The orbital evolution of the comet influenced by non-gravitational acceleration from sublimation \citep{Davidsson2005,Attree2019,Jewitt2020}, as well as the changes in its spin state due to the non-gravitational torque \citep{Keller20151,Attree2024}, are both complex topics, whether from the perspective of forward evolution, or from the perspective of fitting observational data. If the thermophysical model is embedded within a complex physical process as a sub-mechanism, such as the shape evolution \citep{Zhao2021} and the coma formation \citep{Shi2018}, conducting large-scale or high-resolution studies is also challenging.

In our previous work, we addressed similar issues in the thermophysical modelling for asteroids by introducing deep learning techniques \citep{Zhao2024}. We apply a similar concept in this study to develop a deep learning-based approach for the comets that can efficiently solve the thermophysical process and produce the water production rate. 

This paper covers: (i) how the deep learning model can be applied to model the thermal physics of comets; (ii) whether the deep learning model can serve as a unified model for simulating the water activity of the comets in large parameter spaces; (iii) how the deep learning model can be utilized in the analysis of real comets data.

\section{Thermophysical model and water activity}\label{sec2:model}
In this work, we adopt the so called "dust mantle model", in which a dry dust layer is present at the surface of the nucleus over the dust-ice mixture and the water ice sublimation occurs at the boundary between the mantle and the ice front \citep{Hu2017} (and references therein). 

\subsection{Dust mantle model}\label{subsec21:dustm}
The nucleus is regarded as a polyhedron consisting of triangular facets. For each facet, neglecting the changes in thermal properties due to the gas from ice sublimation filling the porosity of the nucleus, the 1D heat conduction equation can be expressed as follows:
\begin{equation}
\label{eqn:diffusion}
\rho C \frac{\partial}{\partial t} T = \kappa \frac{\partial^2}{\partial x^2} T
,\end{equation}
where $\rho$, $C$, $\kappa$ respectively correspond to the density, specific heat capacity and thermal conductivity. They are assumed to be different constants in dry dust mantle and dust-ice mixture, denoted as $\rho_d$, $C_d$, $\kappa_d$ and $\rho_m$, $C_m$, $\kappa_m$. In fact, $\rho_m C_m$, $\kappa_m$ can be expressed in terms of $\rho_d$, $C_d$, $\kappa_d$ and $\rho_i$, $C_i$, $\kappa_i$ of the pure water ice \citep{Davidsson2002,Davidsson2005}:
\begin{equation}
\begin{aligned}
\label{eqn:rhoC_norm}
&\rho_m C_m = (1-f)\rho_d C_d+f\rho_i C_i,\\
&\kappa_m = h((1-f)\kappa_d+f\kappa_i)
\end{aligned}
\end{equation}
where $f$ is the icy area fraction, $h$ is the Hertz factor dependent on the morphological structure of the medium and is generally regarded as a free parameter \citep{Davidsson2002}.

The surface boundary condition of the dust mantle derived from conservation of energy, reads \citep{Kuhrt1994, Lagerros1996}
\begin{equation}
\label{eqn:condition1}
\epsilon \sigma T^4_{x=0}-\kappa_d \frac{\partial}{\partial x} T_{x=0} = (1-A_B) s \psi F_\odot
,\end{equation}
where $\epsilon$ is the emissivity, $\sigma$ is the Stefan–Boltzmann constant. $(1-A_B)s\psi F_\odot$ describes the radiation flux received by the facet, $A_B$ is the Bond albedo, $s$ indicates whether the facet is illuminated by the Sun, considering the horizon shadows and projected shadows, $\psi$ represents the cosine function of the solar elevation angle and $F_\odot$ is the incident solar radiation flux determined by $F_\odot=S_\odot/r^2$, where $r$ is the heliocentric distance of the asteroid and $S_\odot$ is the solar constant.

It is necessary to introduce an additional boundary condition at the interface between the dust mantle and dust-ice mixture, due to the energy consumption caused by water ice sublimation. Assuming the thickness of the dust mantle is $X$, the condition can be expressed as \citep{Kuhrt1994, Hu2017}:
\begin{equation}
\label{eqn:condition2}
-\kappa_d \frac{\partial}{\partial x} T_{x=X^-} = -\kappa_m \frac{\partial}{\partial x} T_{x=X^+}+l Z_{(X)}
,\end{equation}
$l$ is the latent heat of water ice, $Z_{(X)}$ is the mass flux of sublimation determined by \citep{Hu2017}
\begin{equation}
\label{eqn:ZX}
Z_{(X)} = \Psi f Z^{\mathrm{H-K}}(T_{(X)})
,\end{equation}
where
\begin{equation}
\label{eqn:ZHK}
Z^{\mathrm{H-K}}(T) = \alpha(T) P_V(T) \sqrt{\frac{\hat m}{2\pi k_B T}}
,\end{equation}
is the Hertz-Knudsen formula, describing the sublimation flux from the surface of pure solid ice into vacuum \citep{Kossacki1999, Gundlach2011}. $\hat{m}$ is the mass of a water molecules and $k_B$ is the Boltzmann constant. $P_V(T)$ is the saturation vapor pressure. $\alpha(T)$ is the sublimation coefficient, representing the reduction in sublimation flux due to the reattachment of gas molecules to the ice surface after impacting with each other. They are respectively evaluated as \citep{Panale1984, Mauersberger2003, Gundlach2011}
\begin{equation}
\label{eqn:Zdetail}
P_V(T) = a\exp{(-\frac{b}{T})},\;\alpha(T) = c_0+\frac{c_1}{1+\exp{(c_2-c_3/T)}}
,\end{equation}
with constants $a$, $b$, $c_0$, $c_1$, $c_2$, $c_3$ taking the values 3.23$\times 10^{12}$ Pa, 6134.6 K, 0.146, 0.854, 57.78, 11580 K.

The parameter $\Psi$ in Eq. (\ref{eqn:ZX}) represents the reduction factor in the sublimation flux from the bare icy surface, accounting for the permeability of the mantle to gas flow. It is given by the expression $1/(1+pH)$ \citep{Gundlach2011, Hu2017}, where $p$ is a constant and $H=X/d$, with $d$ denoting the diameter of the spherical dust aggregates in the dust mantle \citep{Skorov2012}. We take $p=0.14$ and $d=1$ mm in this work, same as \citet{Hu2017}. $f$ in Eq. (\ref{eqn:ZX}) is used to approximate the reduction in sublimation flux from the pure ice caused by the presence of mixed dust, as the Hertz-Knudsen formula is derived for pure ice \citep{Crifo1997}.

The last boundary condition describes the invariability of temperature at infinitely large depth,
\begin{equation}
\label{eqn:condition3}
\frac{\partial}{\partial x} T_{x \rightarrow \infty} = 0
.\end{equation}

\subsection{Normalization of thermophysical model}\label{subsec22:norm}
Direct application of deep learning methods to the model described above presents several challenges: (1) the parameters have a large dynamic range, leading to the fact that the resulting temperature distribution could span over a wide numerical range. (2) The model is influenced by the rotational velocity of comet, which is not explicitly represented in the established equations. Instead, it is implicitly embedded in the time-dependent function $s \psi$ and the time steps used in the subsequent numerical simulations. Therefore, normalization becomes particularly crucial in this context.

We denote the rotational angular velocity as $\omega$ and introduce the following normalized variables \citep{Lagerros1996}, 
\begin{equation}
\label{eqn:method_norm}
\overline{x}_d = \frac{x}{l_{sd}},\;\overline{x}_m = \frac{x}{l_{sm}},\;\overline{t} = \omega t,\;\overline{T} = \frac{T}{T_e},
\end{equation}
where $l_{sd}$ and $l_{sm}$ are the skin depth of the dust mantle and the dust-ice mixture, $T_e$ is the theoretical maximum surface temperature at the heliocentric distance $r$, called characteristic temperature, given by
\begin{equation}
\label{eqn:detail_norm}
l_{sd} = \sqrt{\frac{\kappa_d}{\rho_d C_d \omega}},\;l_{sm} = \sqrt{\frac{\kappa_m}{\rho_m C_m \omega}},\;T_e = \sqrt[4]{\frac{(1-A_B)F_\odot}{\epsilon \sigma}},
\end{equation}
the heat conduction equation Eq. (\ref{eqn:diffusion}) and three boundary conditions Eq. (\ref{eqn:condition1}), Eq. (\ref{eqn:condition2}), Eq. (\ref{eqn:condition3}) can then be normalized as:
\begin{equation}
\begin{aligned}
\label{eqn:diffusion_condition123_norm}
&\frac{\partial \overline{T}}{\partial \overline{t}} = \frac{\partial^2 \overline{T}}{\partial \overline{x}^2},\\
&\overline{T}^4_{\overline{x}_d=0}-\Phi_d \frac{\partial}{\partial \overline{x}_d} \overline{T}_{\overline{x}_d=0} = E(\overline{t}),\\
&-\Phi_d \frac{\partial}{\partial \overline{x}_d} \overline{T}_{\overline{x}_d=\overline{X}^-} = -\Phi_m \frac{\partial}{\partial \overline{x}_m} \overline{T}_{\overline{x}_m=\overline{X}^+}+l\overline{Z}_{(\overline{X})},\\
&\frac{\partial}{\partial \overline{x}_m} \overline{T}_{\overline{x}_m \rightarrow \infty}=0,
\end{aligned}
\end{equation}
where $\Phi_d$ and $\Phi_m$ are the thermal parameters \citep{Spencer1989} of the dust mantle and dust-ice mixture, defined by
\begin{equation}
\label{eqn:Phi_detail}
\Phi_d = \frac{\sqrt{\rho_d C_d \omega \kappa_d}}{\epsilon \sigma T_e^3},\;\Phi_m = \frac{\sqrt{\rho_m C_m \omega \kappa_m}}{\epsilon \sigma T_e^3}.
\end{equation}
$E(\overline{t})$ is the radiation flux rescaled in the range [0, 2$\pi$] for variables $\overline{t}$ and [0, 1] for $E$. $\overline{Z}_{(\overline{X})} = Z_{(X)}/(1-A_B)F_\odot$ is the sublimation flux after variation.

It is evident that the radiation flux and the derived temperature are successfully rescaled in the range [0, 1], with all physical parameters explicitly represented in the mathematical expression of the equations. This transformation can significantly enhance the efficiency and ease of dataset creation for deep learning. For instance, when studying different comets, the original model for the radiation flux functions leads to varying domains of definition due to differences in rotational velocities. Additionally, the range of values fluctuates with variable parameters such as Bond albedo $A_B$ and heliocentric distance $r$. This variability complicates the construction of a comprehensive set of radiation flux functions. However, the normalized model can effectively address this issue through the parameter extraction.

Excluding the radiation flux, we have identified and summarized all the independent parameters that influence the equations: 
\begin{equation}
\begin{aligned}
\label{eqn:parameters}
&p_1 = (1-A_B)F_\odot,\;p_2 = \epsilon,\;p_3=X,\\
&p_4 = \rho_d C_d,\;p_5 = \omega,\\
&p_6 = \kappa_d,\;p_7 = \kappa_m,\;p_8 = f.
\end{aligned}
\end{equation}
In the subsequent discussions, we treat all these parameters as variable within a physically reasonable range in order to obtain a generalized model.

\subsection{Numerical treatment}\label{subsec23:ns}
The radiation flux needs to be calculated before solving the equations. It is a periodic function about time, determined by the angle between the normal vector of the surface element and the solar direction, incorporating both the horizon and projected shadows. The calculation method is consistent with our previous work for asteroids (secondary radiation and self-heating are not considered in this work), as described in Section 2.1.2 of \citet{Zhao2024}. It is worth noting that the radiation flux used to train the network should be randomly generated, which will be detailed in Section \ref{sec3:deep}, while the illumination and shadow mentioned here are just specific to real-world situations.

The backward finite difference method is utilized to solve the partial differential equations, considering its computational efficiency. First we divide the continuous variables into a set of space-time grids, where the $i$-th grid of depth is denoted by $i\Delta \overline{x}$ and the $j$-th grid of time is denoted by $j\Delta \overline{t}$. As a result, the heat conduction equation is discretized into
\begin{equation}
\label{eqn:diffusion_difference}
\overline{T}_{i,j+1} = (1-2\frac{\Delta \overline{t}}{\Delta \overline{x}^2})\overline{T}_{i,j}+\frac{\Delta \overline{t}}{\Delta \overline{x}^2}(\overline{T}_{i+1,j}+\overline{T}_{i-1,j}).
\end{equation}
The thickness of the dust mantle is much smaller than that of the dust-ice mixture. To improve computational efficiency while ensuring accuracy, different spatial step sizes are applied in two layers, with a smaller step size $\Delta x_d=1$ mm for the dust mantle and a larger one $\Delta x_m=1$ cm for the dust-ice mixture. $\Delta \overline{x}_d$ and $\Delta \overline{x}_m$ are then obtained through dividing $\Delta x_d$ and $\Delta x_m$ by the skin depth $l_{sd}$, $l_{sm}$. The time step is taken as $\Delta \overline{t}=2\pi/1200$. We denote the temperature at the interface between the two layers as $T_{m,j}$, then the three boundary conditions are expressed in the finite difference scheme as follows:
\begin{equation}
\begin{aligned}
\label{eqn:condition_difference}
&\overline{T}_{0,j}^4-\Phi_d \frac{\overline{T}_{1,j}-\overline{T}_{0,j}}{\Delta \overline{x}_d} = E(j \Delta \overline{t}),\\
&-\Phi_d \frac{\overline{T}_{m,j}-\overline{T}_{m-1,j}}{\Delta \overline{x}_d}+\Phi_m \frac{\overline{T}_{m+1,j}-\overline{T}_{m,j}}{\Delta \overline{x}_m} = l \overline{Z}_{(\overline{X})},\\
&\overline{T}_{n,j} = \overline{T}_{n-1,j}.
\end{aligned}
\end{equation}
The surface temperature $T_{0,j}$ and the interface temperature $T_{m,j}$ can be determined from the known $T_{1,j}$, $T_{m-1,j}$, $T_{m+1,j}$ by taking advantage of Newton--Raphson iterative algorithm to solve the boundary conditions.

The overall temperature need to exhibit a stable and static distribution due to the rotation of the comet. After one complete spin of the nucleus, the temperature should be invariant at the subsolar point, which holds at all depths. This static condition is used as a convergence criterion for the iteration. In this work, the criterion is for the mean absolute error of the normalized temperature between iterations to be below $1\times10^{-4}$.

\subsection{Water activity}\label{subsec23:wa}
The water activity of the comet is effectively represented by the sublimation flux in the model. The global total outgassing flux of water from the comet nucleus is the integral of the sublimation flux over the entire surface. In the polyhedral shape model, this can be expressed as follows \citep{Hu2017}:
\begin{equation}
\label{eqn:Ztol}
Z_{\mathrm{total}} = \sum_k{Z_k S_k}
\end{equation}
where $Z_k$ indicates the sublimation flux of $k$-th facet and its area is denoted as $S_k$. Furthermore, we calculate the global water production rate by averaging the global outgassing flux of water over one rotational period, which quantifies the number of water molecules produced per second, given by \citep{Hu2017}
\begin{equation}
\label{Ztolave}
n_Z(t_0) = \frac{1}{\hat{m} P} \int_{t_0}^{t_0+P} Z_{\mathrm{total}} \mathrm{d}t
\end{equation}
where $P$ is the rotational period.

\section{$ThermoONet$}\label{sec3:deep}
In \citet{Zhao2024}, we showed that many factors in a thermophysical model makes the equation-driven deep learning models like physics-informed neural networks (PINNs) \citep{Raissi2019} unsuitable, despite their successful application in solving ordinary differential equations (ODEs) and partial differential equations (PDEs) \citep{Cai2020,Zobeiry2021,He2021,Martin20221,Martin2022,Mathews2023Solving,Laghi2023}. Therefore, for asteroid thermophysical modelling, we adopted deep operator neural network (DeepONet) with strong generalization abilities \citep{LuLu2021,Garg2022,He2023,Branca2024}. Such a network is well suited for handling problems with variable boundary conditions. Following the same consideration, we continue to utilize DeepONet for the development of $ThermoONet$ in this work.

The surface of an asteroid is treated as a layer of dust, whose temperature is influenced primarily by the radiation flux and the thermal parameter (a total of two independent parameters). However, for comets, the distinct physical properties of the dust-ice mixture layer in contrast to the dust mantle, along with the phase transitions of water ice, necessitate a more complex selection of parameters to describe the thermophysical process. Consequently, we identified a total of eight independent parameters (Eq. \ref{eqn:parameters}). In this work, we modified the structure of the neural network to address the high dimensionality of the parameters and furthermore manually optimized the distribution of the training dataset, based on the relative impact of different parameters on the results. The specific methods are introduced in detail below.

\subsection{Architecture of neural network}\label{subsec31:nns}
The principles and structure of DeepONet are in details explained in \citet{LuLu2021}, and we just provide a brief summary here. The core idea is to utilize a neural network to map the different variables from the partial differential equations into a common mathematical space for processing. The network that maps the parameters or functions of the equations is referred to as the branch network, while the network that maps the independent variables is called the trunk network. If we denote the parameters or functions, independent variables as $u$, $y$, these two mapping relationships can be expressed as $u\rightarrow G(u)$ and $y\rightarrow G(u)(y)$, where $G$ is called the operator.

In the thermophysical modelling of cometary systems, the operator mapping process involves the radiation flux function $E(\overline{t})$ and eight characteristic parameters from Equation (\ref{eqn:parameters}), with depth and time serving as independent variables. A DeepONet architecture could theoretically be constructed with branch networks encoding radiation flux function and parameters, trunk networks handling independent variables. The large number of parameters renders the training of an accurate neural network particularly challenging. This necessitates a simplification of the network architecture through problem-specific adaptations. Theoretically the sublimation flux fundamentally depends on the temperature at the dust mantle/dust-ice mixture interface. This physical understanding enables significant architectural simplification by reducing the network output to this subsurface temperature. Consequently, the trunk network becomes redundant and can be eliminated, with the branch network outputs directly providing the temperature predictions. For comprehensive modeling of non-gravitational effects induced by sublimation processes, additional consideration must be given to the surface temperature of the dust mantle, which should be incorporated as another output. This dual-temperature approach enables simultaneous prediction of both the subsurface temperature governing sublimation dynamics and the surface temperature essential for calculating the velocity of the gas (detailed analyses of these effects are provided in \citet{Attree2019} and \citet{Jewitt2020}). 

In the asteroid network, we model the thermal behavior of the asteroid using two separate branch networks: one for encoding the radiation flux function and the other for encoding the thermal parameter \citep{Zhao2024}. The treatment of generating random radiation flux functions can be directly adapted to the comet model (see Section \ref{subsec32:d} for more about the dataset). However, the comet model introduces a larger number of independent parameters exhibiting heterogeneous sensitivity to the solutions of the model, so direct adoption of the parameter encoding scheme for asteroids proves ineffective. A single branch network collectively encoding all parameters can reduce sensitivity to each individual parameter, especially for those parameters significantly influencing the results. Conversely, constructing separate branch networks for individual parameters would result in excessive computational memory allocation and time cost. To resolve this challenge, we propose a parameter grouping method based on sensitivity analysis. Parameters governing the thermal processes are partitioned into two distinct categories, high-sensitivity and moderate-sensitivity parameters. Each category is encoded through dedicated branch network, achieving an optimal balance between parameter sensitivity and computational efficiency. 

\begin{figure*}
\centering
\includegraphics[width=0.75\textwidth]{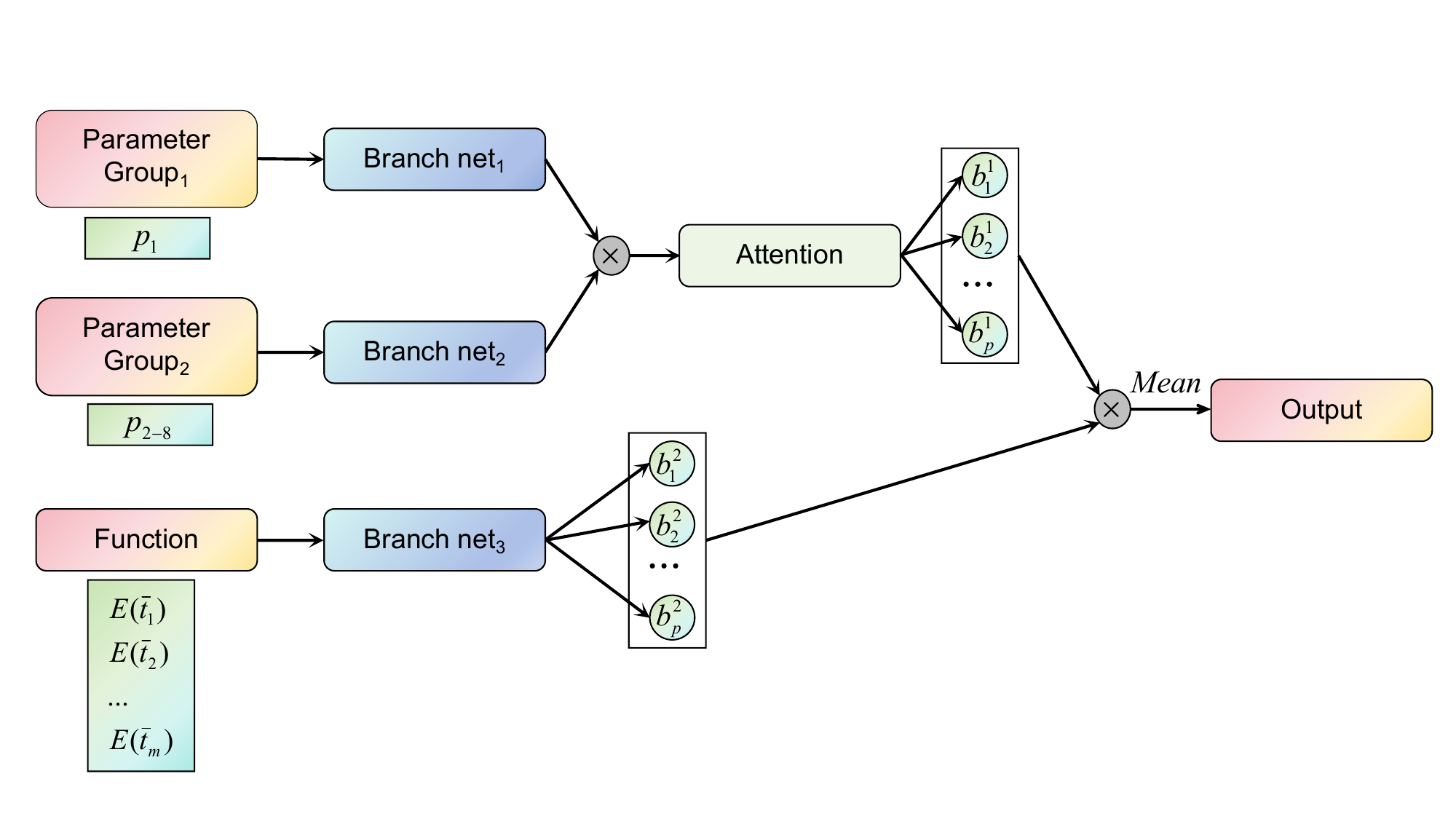}
\caption{Architecture of $ThermoONet$. The network consists of three branch networks with distinct inputs, parameter group $p_1$, $p_{2-8}$ and the radiation flux function $E$, where $E$ is presented as some discrete locations with $m$ elements sampled on $[\overline{t}_1,\overline{t}_2,...,\overline{t}_m]$. A layer incorporating the attention mechanism processes the Hadamard product of the outputs from branch net$_1$ and branch net$_2$. The resulting output from this layer $[b_1^1,b_2^1,...,b_p^1]$ is then element-wise multiplied (Hadamard product) with the output from branch net$_3$ $[b_1^2,b_2^2,...,b_p^2]$. The average of the resulting values is taken as the final output.}
\label{fig:architecture}
\end{figure*}

\begin{figure*}
\centering
\includegraphics[width=0.75\textwidth]{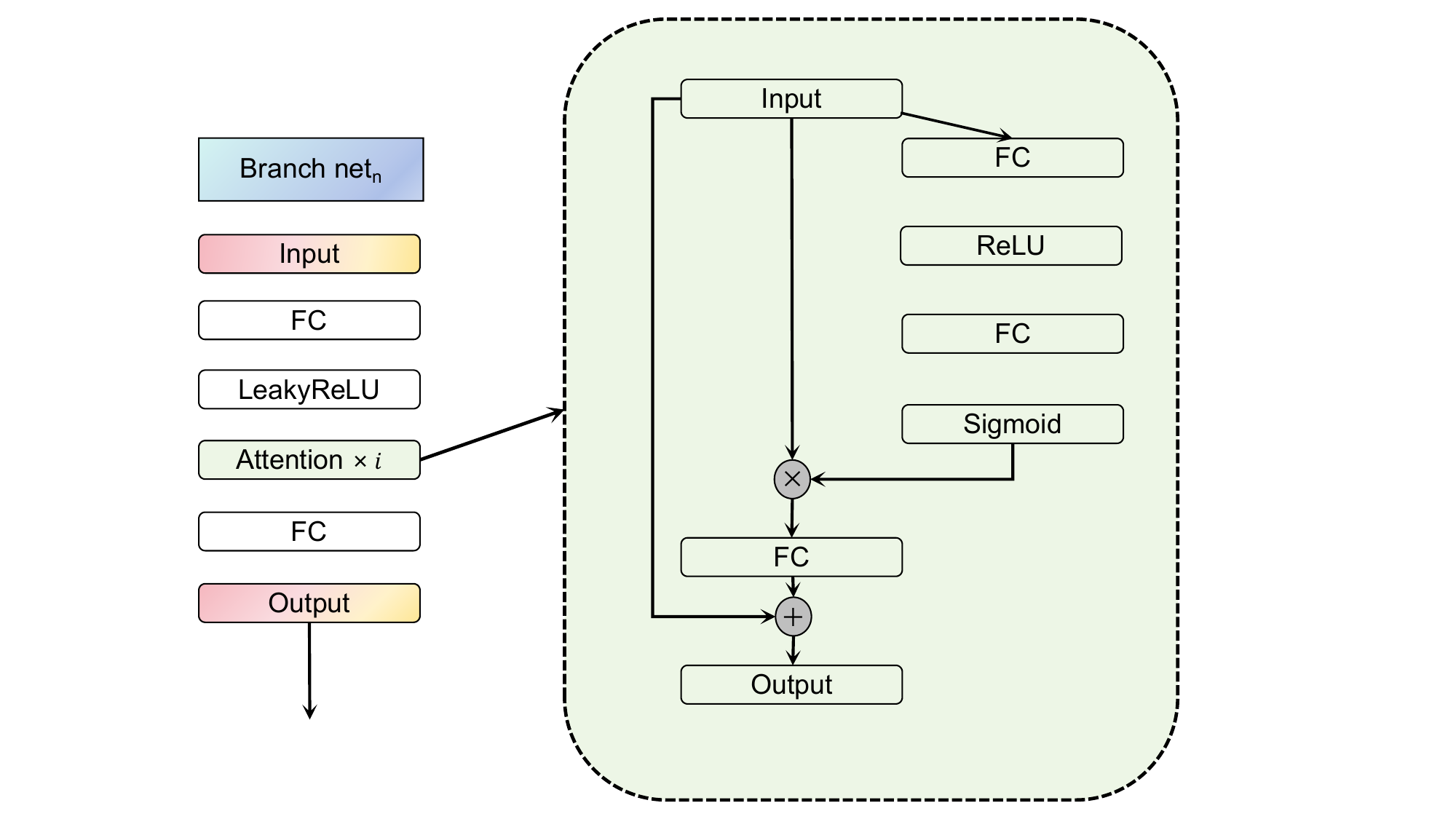}
\caption{Detailed architectures of branch net$_n$ and attention mechanism, where FC is the full connected layer. The input of branch net$_n$ experiences the initial feature extraction through a FC layer and then carries out the feature amplification by attention mechanism, where $i$ represents the layer number. 3, 6, and 6 layers are respectively used in branch networks from net$_1$ to net$_3$.}
\label{fig:detail}
\end{figure*}

The schematic architecture of the neural network adapted for this work is shown in Fig. \ref{fig:architecture}, where the branch networks are presented in Fig. \ref{fig:detail}. Parameter group$_1$ consists of the high-sensitivity parameters $p_1$ which exhibits dominant influence over system behavior (more details are shown in the Appendix \ref{appendix:A}), while parameter group$_2$ consists of the remaining parameters. The neural network encodes these grouped parameters through dedicated branch networks (branch net$_1$ and branch net$_2$), followed by a Hadamard product operation between their respective outputs. This combined tensor undergoes transformation through an attention mechanism layer, designed to amplify the features through adaptive weight allocation \citep{Vaswani2017,Woo2018,Wang2020}. Notably, each branch network incorporates internal attention mechanisms to perform hierarchical feature enhancement. The target subsurface temperature is derived from a averaging operation to the Hadamard product between the amplified features and branch net$_3$ output.

\subsection{Dataset and training}\label{subsec32:d}
The neural network developed in this work focuses on the diurnal water ice activity. Seasonal variations in thermal conditions are reflected through variations in insolation.

As mentioned above, the neural network will be trained with a 9-dimensional parameter space, comprising scalar parameters $p_1$-$p_8$ and one functional input $E$ representing the normalized diurnal insolation curve. To ensure $ThermoONet$ generalizing well across different cometary environments, we should design a training dataset to encompass a broad, diverse set of diurnal radiation flux curves.

For training the asteroid network, we produced a dataset with random pairs of the radiation flux function $E$ and the thermal parameter $\Phi_d$ \citep{Zhao2024}. The corresponding temperatures were calculated by the numerical simulation. Random functions are generated by the mean-zero Gaussian random field (GRF) \citep{LuLu2021}, then rescaled in the range [0,1] through the sigmoid function $1/(1+e^{-(x+d)})$, where $d$ represents the translation distance along the $y$ axis in order to ensure the completeness of the function space \citep{Zhao2024}. For the network of this work, the training process employs a modified stochastic sampling framework, adapted for cometary thermophysical model requirements. The only difference is that all eight parameters undergo simultaneous joint randomization to maintain proper parameter covariance structures. Special treatment is implemented for high-sensitivity parameter $p_1$ through a stratified sampling strategy. $p_1$ is randomly generated within different value ranges, with a similar number in each range. This balanced multi-range generation mechanism, grounded in prior knowledge of the nonlinear dominance of $p_1$, ensures that the network effectively captures the substantial influence of various ranges on the resulting temperature. In the practical training process, we have generated a dataset with 36,000 sets of data.

The loss function is the mean square error (MSE) between the temperature calculated with numerical simulation and the output from the neural network, and the adaptive moment estimation method (Adam) is used as the optimization algorithm. Same as in the training process of the asteroid network in \citet{Zhao2024}, the learning rate decay approach is utilized to mitigate issues related to wide oscillations of the loss function and slow convergence. 

We term the trained neural network $ThermoONet$, a model designed based on the core principles of DeepONet and specifically tailored for cometary thermophysical modelling.

\section{Results}\label{sec4:res}
The cost of computation time and the level of accuracy are used to assess the performance of $ThermoONet$.

\subsection{Computational cost}\label{subsec41:tce}
We compare the computational time consumed by three approaches, $ThermoONet$ with GPU, $ThermoONet$ with CPU, and traditional numerical simulation, for calculating the global ice front temperature distribution of a spherical nucleus with varying number of facets. It is worth mentioning that dataset construction required approximately 6 hours, while the neural network training process was completed within 20 minutes. Once trained, the network is ready for immediate application and does not require further adjustment, so we only analyze the inference time of the network. The $ThermoONet$ calculations were carried out on both GPU and CPU for over 20 times to get an estimate on the uncertainty. The time consumption of the numerical simulation is evaluated by multiplying the average time cost per facet (calculated based on the temperature of a large number of facets) by the total number of facets. All of the computations were performed on a personal computer\footnote{A laptop equipped with AMD Ryzen 9 6900HX with Radeon Graphics and NVIDIA GeForce RTX 3070Ti Laptop GPU} and under the environment of Python 3.8. 

\begin{figure*}
\centering
\includegraphics[width=0.8\textwidth]{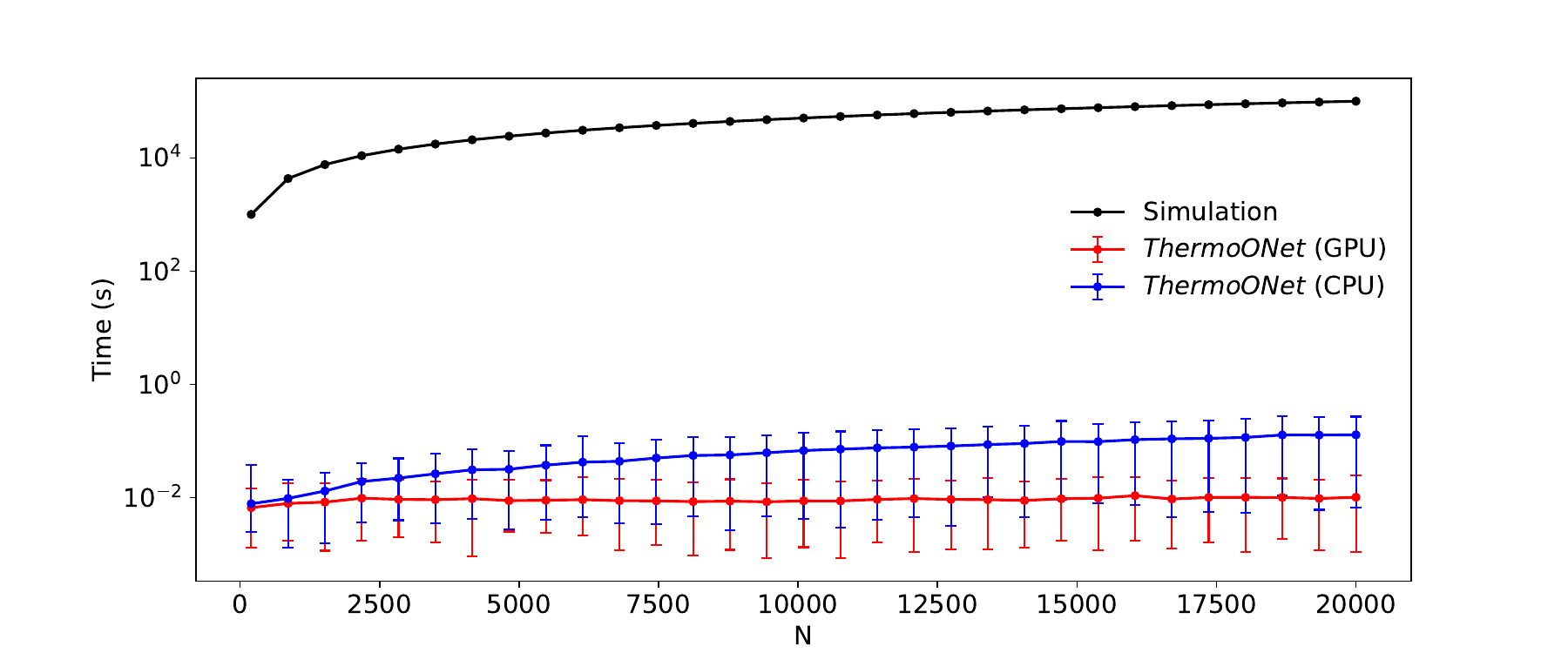}
\caption{Comparison of computational cost between $ThermoONet$ and traditional numerical simulation for the shape models with varying number of facets. The error bars represent the maximum and minimum time costs in the repeated calculation for over 20 times.}
\label{fig:timecost}
\end{figure*}

Results show that (Fig. \ref{fig:timecost}):
\begin{itemize}
\item[$\bullet$] Calculation using $ThermoONet$ is approximately six orders of magnitude faster than the numerical simulation.
\item[$\bullet$] Computational time costs by numerical and CPU-based $ThermoONet$ increase almost linearly with number of facets, while the computation cost by GPU-based $ThermoONet$ stays almost constant against increasing number of facets.
\end{itemize}

When we need to calculate a large number of global temperature distributions, it is worth noting that the vectorized operations can be further used to ensure that the computation time does not increase linearly with the number of facets. 

\subsection{Accuracy assessment}\label{subsec42:ea}
We first use a spherical comet to assess the accuracy of temperature predictions with varying obliquity $\beta$ (the angle between the spin axis and the normal vector of the orbital plane) and heliocentric distance $r$. We also investigate how errors change with different key model parameters such as the thickness of the dust mantle $X$ and icy area fraction $f$, where fixed parameters are listed in Table \ref{tab:pa0}. We measure the mean absolute percentage error (MAPE) of temperature between the neural network and numerical simulation, denoted as MAPE ($T$). 

\begin{figure*}
\centering
\includegraphics[width=0.8\textwidth]{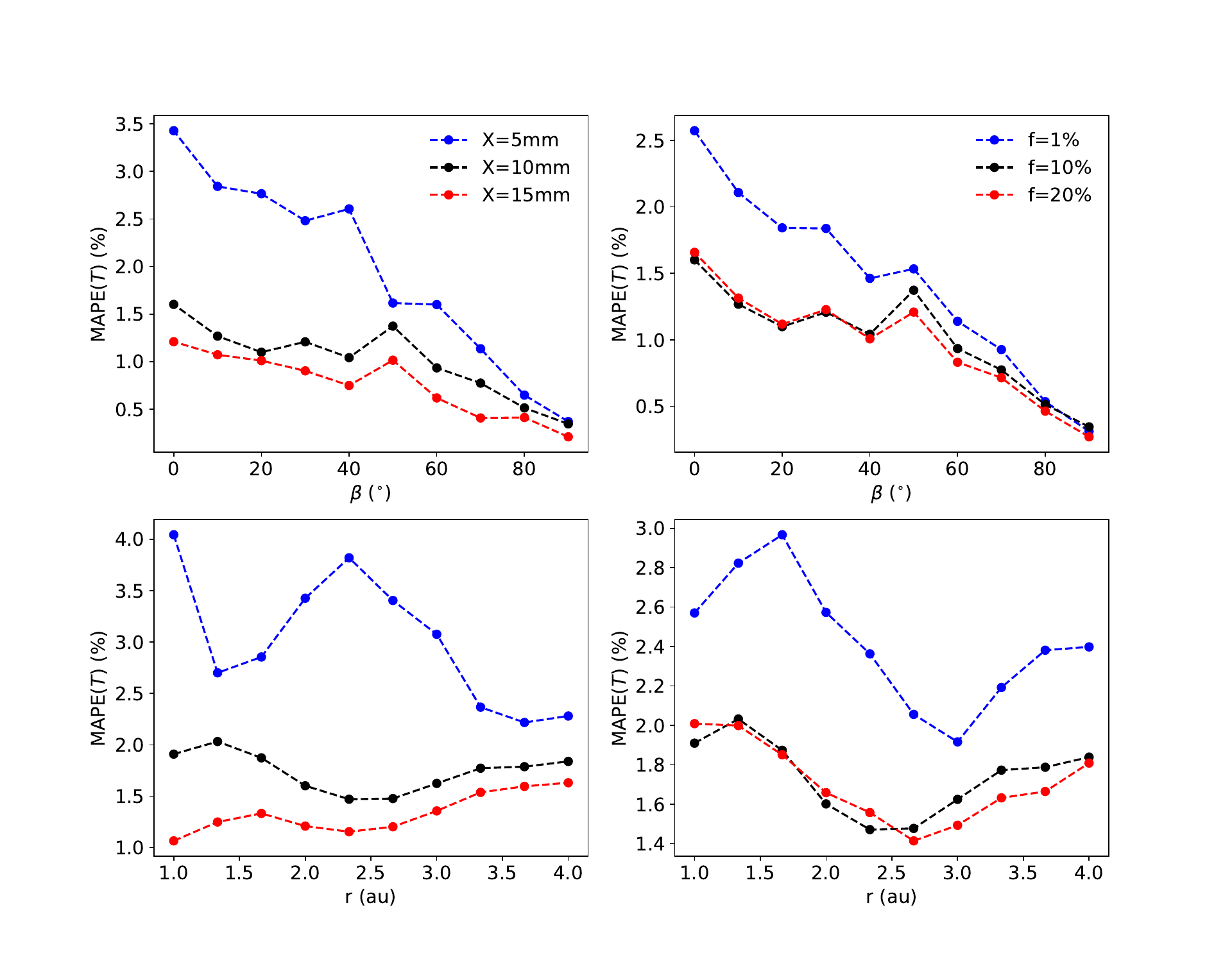}
\caption{Differences in the output subsurface temperature between $ThermoONet$ and numerical simulation as functions of obliquity $\beta$ and heliocentric distance $r$. The left panels present the error functions for different thicknesses of dust mantle, with the blue, black, and red lines respectively representing 5 mm, 10 mm, and 15 mm respectively. The right panels illustrate the error functions for varying icy area fractions, with the blue, black, and red lines respectively corresponding to 1$\%$, 10$\%$, and 20$\%$. The panels fix $r=2$ au for the top panels and $\beta=0^{\circ}$ for the bottom panels, $f=10\%$ for the left panels and $X=10$ mm for the right panels. For the settings of other parameters, please refer to Table \ref{tab:pa0}.}
\label{fig:error1}
\end{figure*}

Fig. \ref{fig:error1} shows the results. The overall error in temperature fluctuates as the radiation flux function or parameters change. With the inclusion of more parameters, the error inevitably increases compared to that of the deep learning model for asteroids \citep{Zhao2024}. However, it is still maintained at a reasonably low level, with maximum below 4$\%$ and an average around 2$\%$. We notice that the error does not show significant correlation with varying heliocentric distances, which confirms the effectiveness of the specially designed training dataset. 

\begin{table}[]
\footnotesize
\centering
\caption{Fixed parameters in the error analysis.}
\setlength{\tabcolsep}{4 mm}{
\begin{tabular*}{0.4\textwidth}{@{\extracolsep{\fill}}lc@{\extracolsep{\fill}}}
    \hline
        {$\epsilon$} & 1 \\ \hline
        {$\rho_i$ ($\mathrm{kg\;m^{-3}}$)} & 920 \\ \hline
        {$C_i$ ($\mathrm{J\;kg^{-1}\;K^{-1}}$)} & 2100 \\ \hline
        {$\rho_d$ ($\mathrm{kg\;m^{-3}}$)} & 500 \\ \hline
        {$C_d$ ($\mathrm{J\;kg^{-1}\;K^{-1}}$)} & 1000 \\ \hline
        {$\omega$ ($\mathrm{rad\;s^{-1}}$)} & 1.4$\times\;10^{-4}$  \\ \hline
        {$\kappa_d$ ($\mathrm{W\;m^{-1}\;K^{-1}}$)} & 2$\times\;10^{-3}$ \\ \hline
        {$\kappa_m$ ($\mathrm{W\;m^{-1}\;K^{-1}}$)} & 2.4$\times\;10^{-3}$ \\ \hline
\end{tabular*}}
\label{tab:pa0}
\end{table} 

\begin{figure*}
\centering
\includegraphics[width=0.84\textwidth]{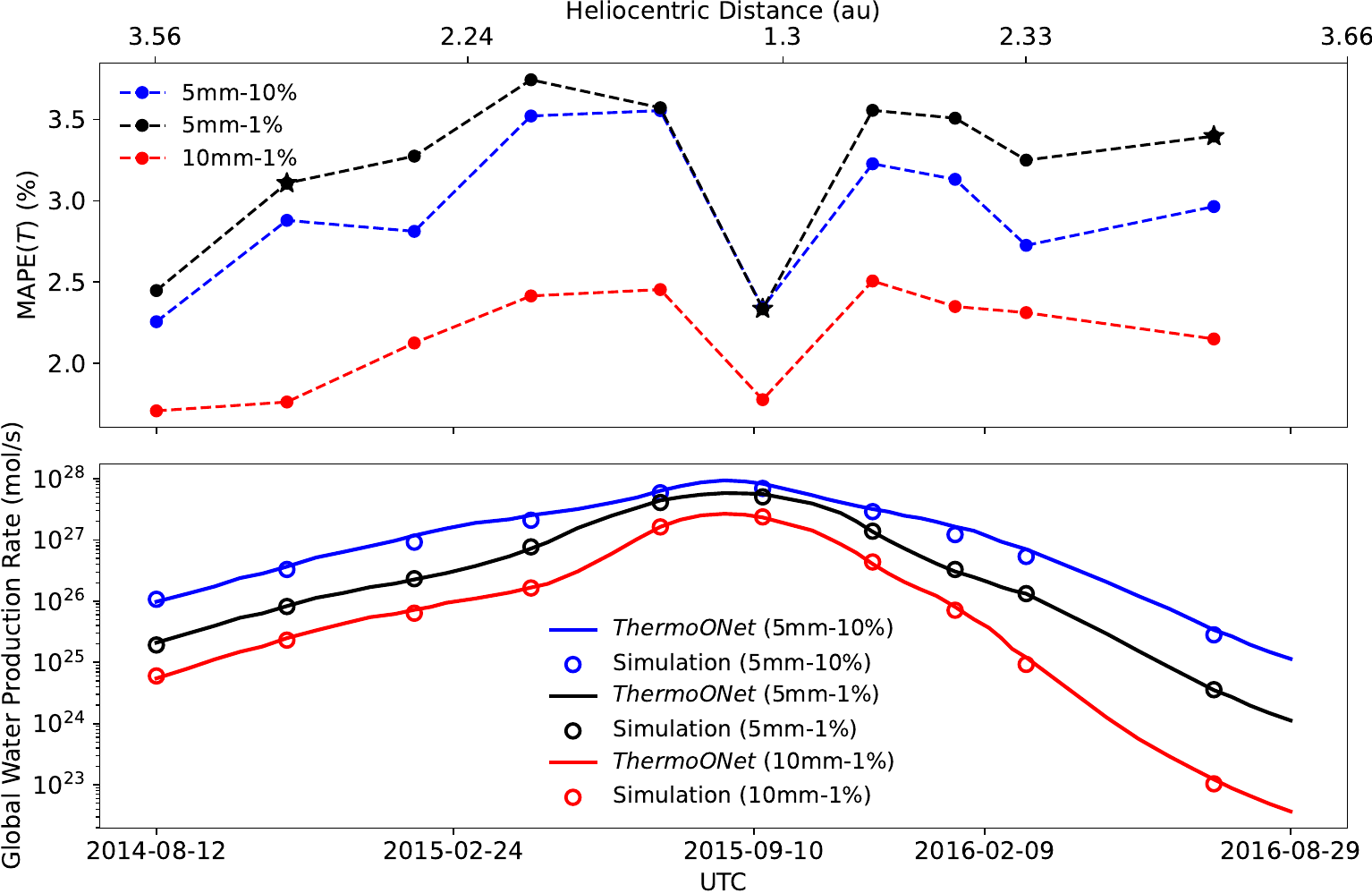}
\caption{Differences in global water production rate derived from $ThermoONet$ and the numerical simulation at different orbital phases of comet 67P. The top panel is the temperature calculation errors, where the black line was derived with $X=5$ mm, $f=1\%$, and the red line with $X=10$ mm, $f=1\%$, the blue line with $X=5$ mm, $f=10\%$. The asterisks denote specific examples of temperature distributions shown in Fig. \ref{fig:67P_detail}. The bottom panel is the water production rate, where the real lines stand for the results from $ThermoONet$, the circles are these from the numerical solutions, with the same parameters as in the top one.}
\label{fig:error2}
\end{figure*}

\begin{figure*}
\centering
\includegraphics[width=0.8\textwidth]{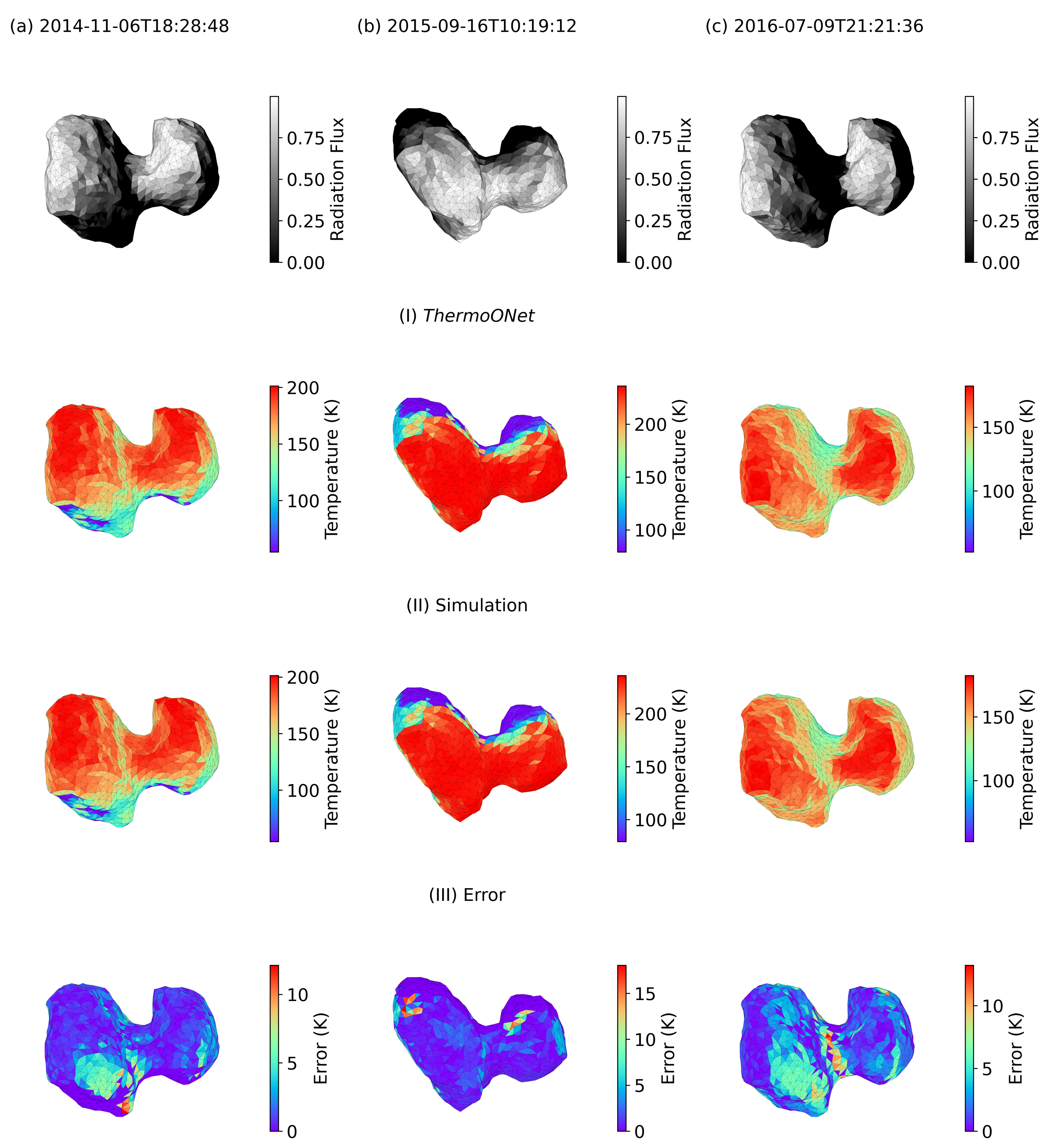}
\caption{The subsurface temperature distribution of comet 67P modelled by $ThermoONet$ and the numerical simulation at three specific positions marked by asterisks in the top panel of Fig. \ref{fig:error2}. The top row shows the normalized radiation flux, and the second to fourth rows show the results calculated with $ThermoONet$, the numerical simulation, and the differences between them.}
\label{fig:67P_detail}
\end{figure*}

As shown in Eq. \ref{eqn:ZX}-\ref{eqn:Zdetail}, the sublimation flux of water ice exhibits an exponential relationship with the temperature, meaning that any error in temperature has an even more significant impact on the calculated sublimation flux. Moreover, given that many cometary nucleus show a concave shape, the projected shadows between facets complicate the radiation flux function, which presents a great challenge for the generalization of the neural network, even more severe than in the case of asteroid \citep{Zhao2024}. In order to validate the generalization capability of $ThermoONet$, it is applied to a real comet with irregular shape. In practice, we take comet 67P as the example and simulate its temperature with a shape model of 2868 facets \citep{Jorda2016}. Ephemerides of 67P is retrieved using Rosetta SPICE kernels \citep{Acton1996,ESS}. We evaluate both the temperature and water production rate at different orbital phases by comparing them with the results from numerical solutions (see Fig. \ref{fig:error2}). We also examine the temperature distributions at three specific orbital locations before, near and after the perihelion (Fig. \ref{fig:67P_detail}). Other parameters are the same as in Table \ref{tab:pa0}. 

The average temperature errors are about 2.5$\%$, and the global water production rates from two methods are showing the same trend. The average relative errors are respectively 5.04$\%$ for ($X=5$ mm, $f=1\%$), 19.25$\%$ for ($X=5$ mm, $f=10\%$), and 10.26$\%$ for ($X=10$ mm, $f=1\%$). The patterns of error distribution shown in Fig. \ref{fig:error2} suggest that the temperature error is the smallest in the high-temperature region, typically below 3K. The primary source of error lies in the mid-temperature regions. However, only a few facets exhibit errors exceeding 10K.

\section{Application to the interpretation of water production rate}\label{sec5:app}
Interpreting the water production rate as a function of the heliocentric distance using thermophysical models has been an intriguing topic in cometary science as it helps us understand the nature of cometary activity \citep{Keller2015,Hu2017,Attree2019,Marshall2019,Skorov2020}. It is a time-consuming process to calculate the water production rate curve (hereafter water curve) using traditional numerical simulation, which requires computations of global temperature distribution at different orbital phases and it needs to be recalculated whenever the parameters are adjusted. Due to the large number of physical parameters involved, previous studies often fix centain parameters to some typical values, leaving only a few to be adjustable. When fitting the observational data, it is difficult to take advantage of the automated algorithms to extract parameters from the observational data, thus, the parameters are typically determined through empirical trial-and-error testing.

Generating water curves with $ThermoONet$ is efficient, hence enabling the application of global optimization algorithms, such as Genetic Algorithm (GA), Particle Swarm Optimization (PSO), Simulated Annealing (SA) and so on, to find the best-fitting parameters. 

To do that, we implement $ThermoONet$ to calculate the water curve in three steps:
\begin{itemize}
\item[$\bullet$] a lookup table is generated by calculating illumination and shadows under varying solar positions, incorporating the shape and rotational state of the comet. This is achieved through Möller-Trumbore algorithms and interpolation algorithms.
\item[$\bullet$] eight scalar parameters are computed based on the physical properties of the comet and orbital ephemerides, the radiation flux for each facet then is retrieved from the precomputed lookup table.
\item[$\bullet$] the derived 9-dimensional parameters serve as input to $ThermoONet$, which outputs subsurface temperature at the ice front of each facet. The results subsequently drive the calculation of water ice sublimation rates and associated activity levels.
\end{itemize}

\subsection{Fitting the water curve of 67P}\label{subsec51:67P}
We combine $ThermoONet$ with SA to fit the observed water curve of comet 67P from \citet{Lauter2020}, the shape model with 23788 facets is used \citep{Jorda2016}.

\begin{figure*}
\centering
\includegraphics[width=0.72\textwidth]{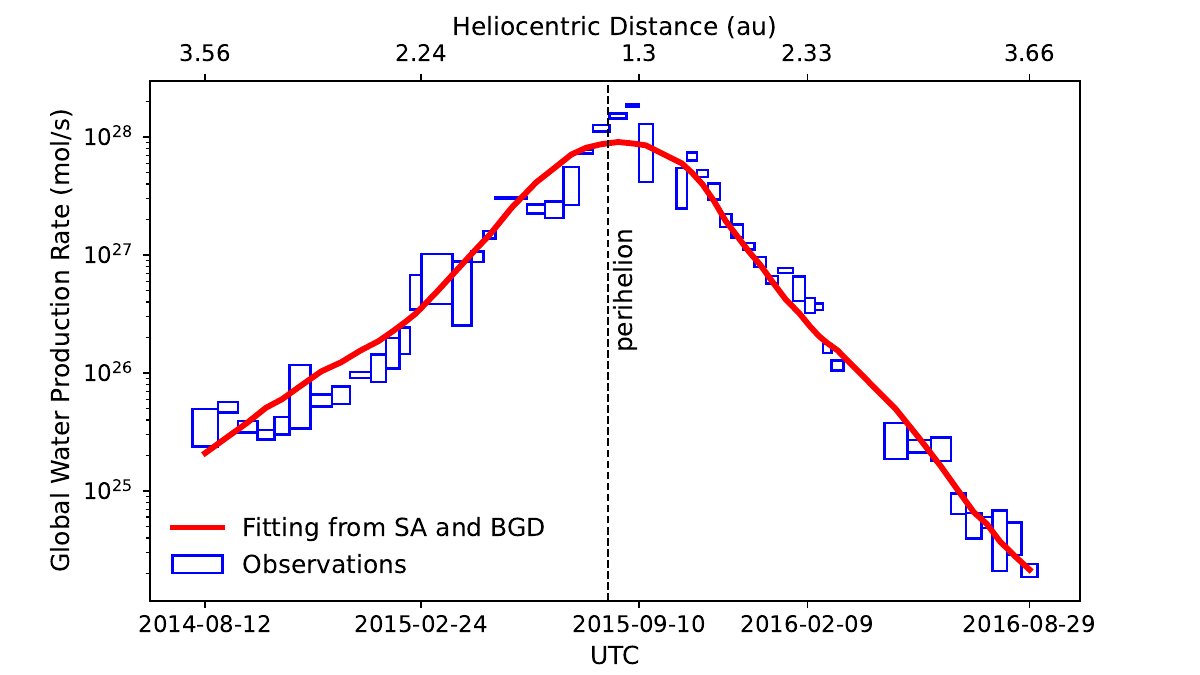}
\caption{Fitting the global water production rate of 67P as a function of time. The blue rectangles represent the observation data from \citet{Lauter2020}, and the red line corresponds to the best fit curve achieved by combining $ThermoONet$ with the optimization algorithm. The black dashed line indicates the epoch of the perihelion. A detailed description of the observation data and sources thereof can be found in \citet{Lauter2020}.}
\label{fig:fitting}
\end{figure*}

We chose five parameters to be fitted, including the thickness of the dust mantle $X$, the icy area fraction $f$, the product between the density of the dust mantle $\rho_d$ and the specific heat capacity $C_d$, the heat conductivity of the dust mantle $\kappa_d$ and the dust-ice mixture $\kappa_m$. The initial values of the five parameters are selected as in Table \ref{tab:pa1} and the corresponding water curve is generated by applying $ThermoONet$ at every observation time. The loss function is MAPE between the predicted results and the average observed data, and SA is applied to search for the preliminary optimal parameter combination that minimizes MAPE. Finally, Batch Gradient Descent (BGD) is utilized to refine the parameters, converging to a solution that further reduces the loss function. The best fitting water curve is shown in Fig. \ref{fig:fitting} and the optimized parameter combination is shown in Table \ref{tab:pa1}.

\begin{table}[]
\footnotesize
\centering
\caption{Initial values and the results derived from SA and BGD for the variable parameters in fitting the water curve of 67P.}
\setlength{\tabcolsep}{4 mm}{
\begin{tabular*}{0.4\textwidth}{@{\extracolsep{\fill}}lc@{\extracolsep{\fill}}c@{\extracolsep{\fill}}}
    \hline
        {Parameters} & Initial values & Fitting results \\ \hline
        {$X$ (mm)} & 5 & 3.61 \\ \hline
        {$f$ ($\%$)} & 5 & 0.644 \\ \hline
        {$\kappa_d$ ($\mathrm{W\;m^{-1}\;K^{-1}}$)} & 1$\times\;10^{-3}$ & 2.18$\times\;10^{-3}$ \\ \hline
        {$\kappa_m$ ($\mathrm{W\;m^{-1}\;K^{-1}}$)} & 1$\times\;10^{-3}$ & 2.20$\times\;10^{-3}$ \\ \hline
        {$\rho_d C_d$ ($\mathrm{J\;m^{-3}\;K^{-1}}$)} & 2$\times\;10^{5}$ & 5.68$\times\;10^{5}$ \\ \hline
\end{tabular*}}
\label{tab:pa1}
\end{table} 

The overall fitting performance is satisfactory with MAPE = 37.08$\%$, capturing the asymmetrical trend before and after perihelion. A slight underestimation is observed near the perihelion, which is consistent with the finding in \citet{Hu2017}. This phenomenon may be explained by several factors, including the varying levels of activity across different regions of 67P \citep{Attree2019}, the gradual exposure of water ice as the comet approaches perihelion \citep{Fornasier2016,Filacchione2020}, and the temperature-dependent changes in thermal conductivity \citep{Skorov2020}. It is worth mentioning that the advantage of $ThermoONet$ in computation efficiency could be of help in assessing aforementioned mechanisms.

\subsection{Fitting the water curve of 21P}\label{subsec52:21P}
There is a growing database of water curves of different types of comets retrieved using observations by solar observatories \citet{Combi2019,Combi2021b,Combi2023}. Most of these comets still lack well-defined parameters such as size, shape, rotation characteristics, and thermophysical parameters. It could prove difficult for traditional numerical simulations to fit these data by searching a large parameter space. However, by combining trial-and-error testing with $ThermoONet$, exploring a large parameter space becomes feasible. We take comet 21P as an example to demonstrate the application of the trial-and-error testing. The water curve of 21P is significantly different from that of 67P, remaining essentially flat before perihelion and exhibiting an obvious decline after perihelion (see more from \citet{Combi2021a}). A preliminary consideration is that the solar illumination area of 21P gradually decreases during the observation period, which could be the combined effect of its shape and the orientation of its spin axis, which was shown in \citet{Marshall2019}. We can iteratively adjust the parameters based on trial-and-error results by taking advantage of $ThermoONet$.

\begin{figure*}
\centering
\includegraphics[width=0.72\textwidth]{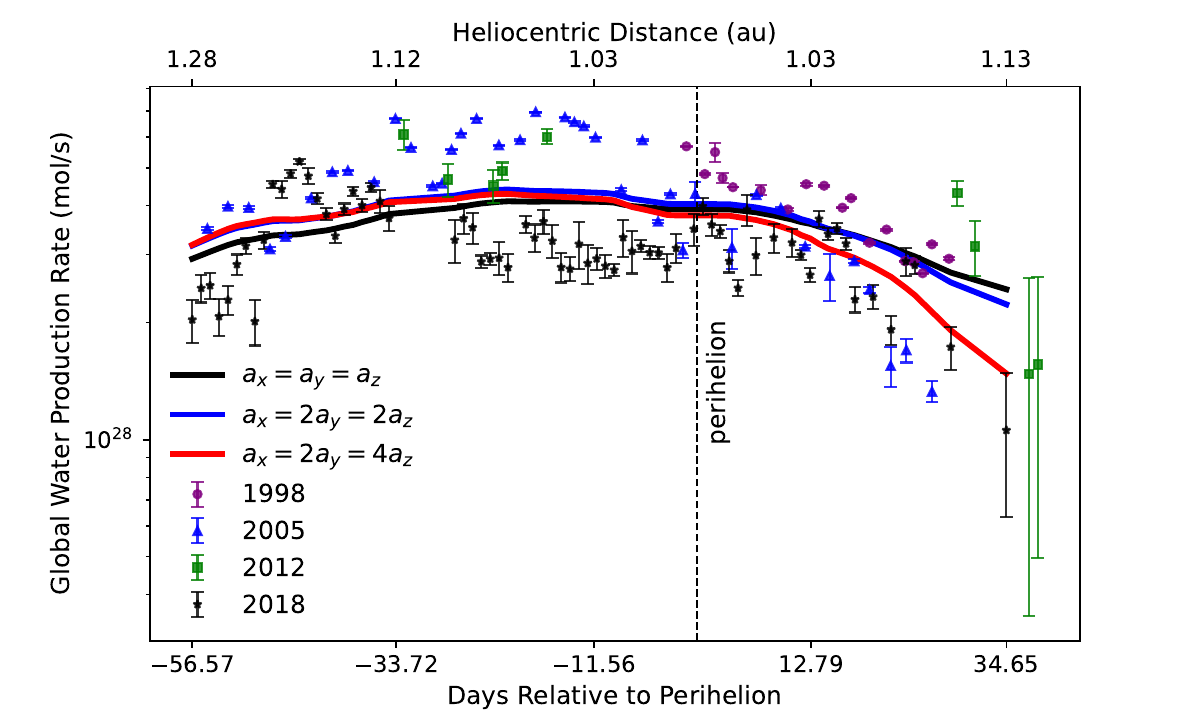}
\caption{Fitting the global water production rate of 21P as a function of days from perihelion by applying $ThermoONet$ to different shapes. The black line corresponds to a sphere model, the blue line is a ellipsoid model with $a_x=2a_y=2a_z$ and the red line is a ellipsoid with $a_x=2a_y=4a_z$. Other parameters are as stated in Table \ref{tab:pa2}. The scatter points with different colors and shapes represent observational data at different apparitions \citep{Combi2019}, where purple circles from 1998, blue triangles from 2005, green squares from 2012, and black asterisks from 2018. The detailed description about the observation data and sources thereof can be found in \citet{Combi2019}.}
\label{fig:fitting_21P}
\end{figure*}

Fig. \ref{fig:fitting_21P} shows preliminary fitting results. The fixed parameters are $\epsilon=1$, $\omega=1.74\times10^{-4}\;\mathrm{rad\;s^{-1}}$ (the rotation period is uncertain with a wide range from 9.5 to 19 h \citep{Leibowitz1986,Moulane2020}). Parameters to be fitted are shown in Table \ref{tab:pa2}. The test shapes are respectively sphere with $a_x=a_y=a_z=3.5$ km, biaxial ellipsoid with $a_x=2a_y=2a_z=5.2$ km and triaxial ellipsoid $a_x=2a_y=4a_z=5.2$ km. As shown in Fig. \ref{fig:fitting_21P}, the evolutionary trend of the triaxial ellipsoid better matches the observational data, confirming our previous hypothesis regarding the solar illumination area, meaning that the farther the comet is before the perihelion, the larger the region with higher temperature. 

\begin{table}[]
\footnotesize
\centering
\caption{Parameters for fitting the water curve of 21P.}
\setlength{\tabcolsep}{4 mm}{
\begin{tabular*}{0.4\textwidth}{@{\extracolsep{\fill}}lc@{\extracolsep{\fill}}}
    \hline
        {$X$ (mm)} & 5 \\ \hline
        {$f$ ($\%$)} & 6 \\ \hline
        {$\kappa_d$ ($\mathrm{W\;m^{-1}\;K^{-1}}$)} & 2$\times\;10^{-3}$ \\ \hline
        {$\kappa_m$ ($\mathrm{W\;m^{-1}\;K^{-1}}$)} & 2.025$\times\;10^{-3}$ \\ \hline
        {$\rho_d C_d$ ($\mathrm{J\;m^{-3}\;K^{-1}}$)} & 5.17$\times\;10^{5}$ \\ \hline
        {Spin axis orientation ($^{\circ}$)} & (RA=-5, Dec=-40) \\ \hline
\end{tabular*}}
\label{tab:pa2}
\end{table} 

\section{Conclusion}\label{sec6:con}
In this work, we developed $ThermoONet$, a DeepONet-based neural network for modelling cometary activity. Due to the high-dimensional parameter space in cometary thermophysical model, we tailored the architecture of DeepONet by removing the trunk network and trained it specifically for the target temperature. With solar flux as input, $ThermoONet$ predicts the subsurface temperatures and water production rate. By taking advantage of the generalization ability of DeepONet, $ThermoONet$ could be used in modelling comets with varying physical parameters and shapes.

We conducted a comprehensive evaluation of the network's performance, including its computation time, accuracy, and generalization ability. The results show that $ThermoONet$ deployed on GPU has the highest efficiency, and generates the global sublimation flux approximately six orders of magnitude faster than using numerical simulation. The accuracy of $ThermoONet$ varies with parameters, with an average of around 2$\%$. $ThermoONet$ can be successfully applied to irregular comets as exemplified by comet 67P, with the water production rate curve closely matching the numerical simulation, demonstrating strong generalization ability and practicability.

We tested the capability of $ThermoONet$ in computationally demanding tasks by fitting global water production rate curves of comets using global optimization algorithms. By applying Simulated Annealing, we successfully fit the observed water production rate curve of comet 67P derived with Rosetta mission data, and obtained a set of optimized physical parameters of 67P. Additionally, we fit the water curve of comet 21P to demonstrate the capability of combining $ThermoONet$ with trial-and-error testing. Although limited knowledge is available for 21P, we managed to reconstruct the unique trend observed in its water production rate by adjusting the shape, rotation, and thermal parameters.

We anticipate a broad range of application for $ThermoONet$ in various research topics concerning small body activities. In addition to analyzing large-scale and high-resolution data analysis, it can also serve as an efficient model to be embedded into more complex physical systems.

\appendix
\section{Parameter sensitivity}\label{appendix:A}
\begin{figure*}
\centering
\includegraphics[width=0.9\textwidth]{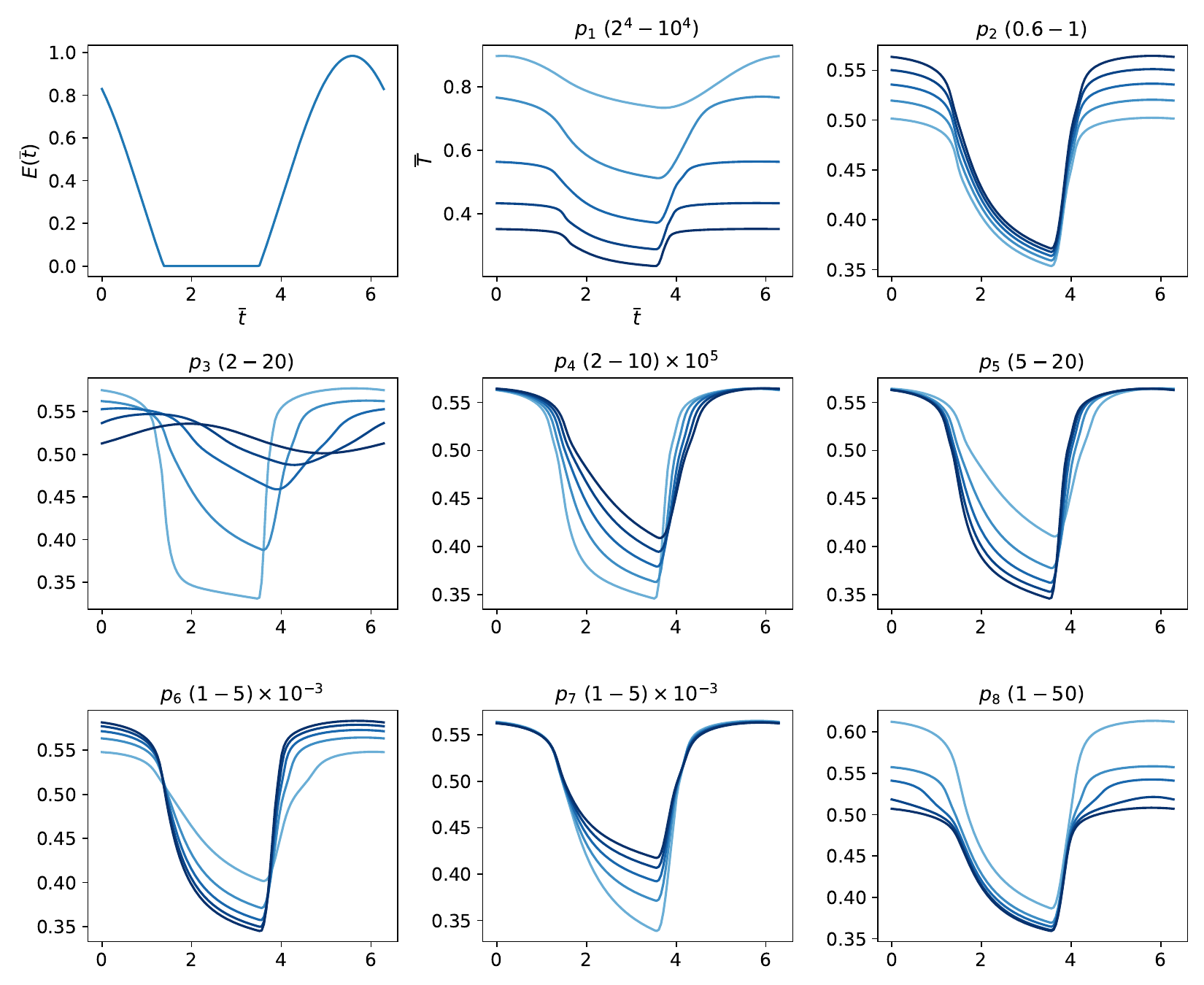}
\caption{Comparative analysis of parameter sensitivity. The first panel depicts the selected example radiation flux, followed by eight panels illustrating different temperature curves under varying values of the eight distinct parameters. The titles of these panels show the variation ranges with units as the same as Table \ref{tab:att11}, and the curves with darker hues correspond to larger values of parameters.}
\label{fig:att12}
\end{figure*}

We conduct the parameter sensitivity analysis using the controlled variable approach with a fixed radiation flux as an example. When modifying a single parameter, the remaining parameters maintain their reference values specified in Table \ref{tab:att11}. Here the rotational angular velocity $p_5$ is represented as the rotation period. Fig. \ref{fig:att12} shows the resultant variations in temperature curves induced by alterations of different parameters. $p_1$ primarily influences the value range of temperature distributions, whereas the remaining parameters affect the shape of temperature curves within this established range.

\begin{table}[]
\footnotesize
\centering
\caption{Parameters and their reference values used in the parameter sensitivity analysis.}
\setlength{\tabcolsep}{4 mm}{
\begin{tabular*}{0.4\textwidth}{@{\extracolsep{\fill}}lc@{\extracolsep{\fill}}}
    \hline
        {$p_1$ ($\mathrm{W\;m^{-2}}$)} & $6^4$ \\ \hline
        {$p_2$} & 1 \\ \hline
        {$p_3$ (mm)} & 5 \\ \hline
        {$p_4$ ($\mathrm{J\;m^{-3}\;K^{-1}}$)} & $5\times10^{5}$ \\ \hline
        {$p_5$ (h)} & 10  \\ \hline
        {$p_6$ ($\mathrm{W\;m^{-1}\;K^{-1}}$)} & 2$\times\;10^{-3}$ \\ \hline
        {$p_7$ ($\mathrm{W\;m^{-1}\;K^{-1}}$)} & 2$\times\;10^{-3}$ \\ \hline
        {$p_8$ ($\%$)} & 10 \\ \hline
\end{tabular*}}
\label{tab:att11}
\end{table} 

\begin{acknowledgements}
The authors thank Prof. Groussin for the insightful review comments that helped to significantly improve the manuscript. The authors thank Dr. Xuanyu Hu for extensive discussions on the thermophysical modelling for comets. XS thanks members of the ISSI International Team "Timing and Processes of Planetesimal Formation and Evolution" for inspirational discussions. This work is financially supported by the National Natural Science Foundation of China (No. 12233003 and 12073011).
\end{acknowledgements}
\bibliographystyle{aa}
\bibliography{references}
\end{document}